\documentclass[final,5p,times,twocolumn]{elsarticle}

\usepackage{amssymb}
\usepackage{hyperref}
\usepackage[font=footnotesize,labelfont=bf]{subcaption}

\usepackage{lineno}

\journal{Nucl. Instr. Meth. A}

\begin{document}

\begin{frontmatter}



\title{Novel photon detectors}


\author[fmful,jsi]{Peter Kri\v zan}

\affiliation[fmful]{organization={Faculty of Mathematics and Physics, University of Ljubljana},
            city={Ljubljana},
            country={Slovenia}}
            
\affiliation[jsi]{organization={Jožef Stefan Institute},
            city={Ljubljana},
            country={Slovenia}}

\begin{abstract}
The paper reviews recent progress in photodetectors, discussing vacuum-based detectors, semiconductor sensors, and gas-based detectors. The emphasis in this review is on the detection of low light levels, enhanced timing resolution, and spectral range of photon detectors, as well as the development of photosensors for extreme conditions, for operation in cryogenic and high radiation-level environments.

\end{abstract}




\end{frontmatter}


\section{Introduction}
\label{intro}

Photon detectors are at the heart of most experiments in particle physics. Moreover, they are also finding application in other scientific fields like solid-state and soft-mater physics, chemistry, and biology, and are ubiquitous in society in general. In particle physics, we are encountering new operational environments where we need to detect light, in particular at low light levels. Therefore, advancements are needed in existing technology, as well as transformative, novel ideas to meet the demanding requirements. Two main lines of R\&D are being pursued, an enhanced timing resolution and spectral range of photon detectors, as well as the development of photosensors for extreme environments; these research and development areas have also been identified by the recently published ECFA Detector R\&D Roadmap~\cite{roadmap}.

This review discusses low-light-level photosensors with some emphasis on the two main R\&D areas discussed above. We will review vacuum-based detectors, semiconductor sensors, and gas-based detectors in detecting low light levels. We will discuss their timing properties, response to high radiation levels, and mitigation of radiation damage effects. We will finish with a summary and outlook.

\section{Vacuum-based photodetectors}
\label{vbd}

\begin{figure}[h!]
\centering
\begin{subfigure}[h!]{\linewidth}
    \centering
     \includegraphics[width=0.85\columnwidth]{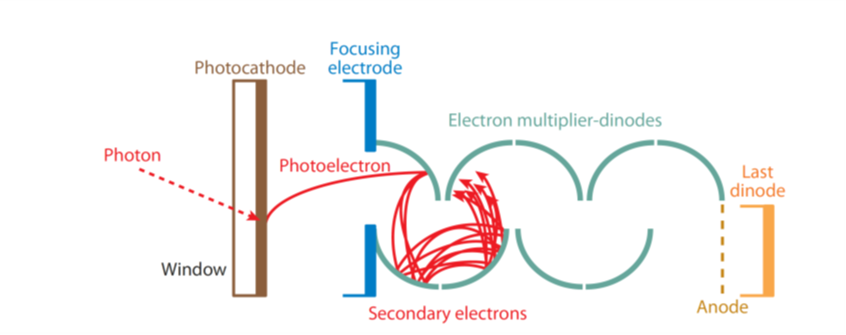}
\end{subfigure}
\caption{Vacuum-based photon detector: principle of operation of a photomutiplier tube~\cite{skpk-review}.}
\label{fig:pmt}
\end{figure}

A vacuum-based photon detector (Fig.~\ref{fig:pmt}) is comprised of 
a photocathode where the conversion of a photon to a photoelectron is carried out, a photoelectron collection and multiplication system made of dynodes, microchannel plates, or a combination of high electric field and a silicon sensor, and finally of a signal collection electrode (anode).


%
\begin{figure}[h!]
\centering
\begin{subfigure}[h!]{\linewidth}
    \centering
     \includegraphics[width=0.85\columnwidth]{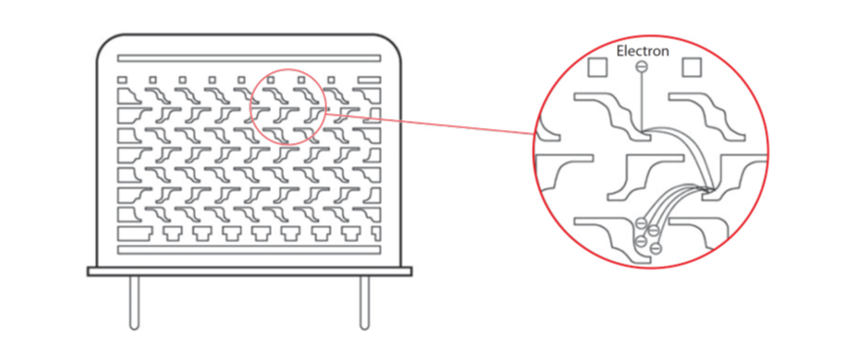}
\end{subfigure}
\caption{Multianode photomultiplier tube: principle of operation~\cite{mapmt-hpk}.}
\label{fig:mapmt}
\end{figure}

\subsection{Multianode photomultiplier tubes}

Multianode photomultiplier tube (MA-PMTs) (Fig.~\ref{fig:mapmt}) is a multi-channel version of a PMT with metal channel dynodes. MA-PMTs usually employ bialkali photocathodes, have a gain of about $\approx 10^6$, and an operating voltage of around 1~kV. They are characterized by excellent performance, excellent single photon detection efficiency, very low noise, and low cross-talk making it the best choice for a position-sensitive photon detector for single photons to cover large areas with no magnetic field (Fig.~\ref{fig:mapmt}).
\begin{figure}[h!]
\centering
\begin{subfigure}[h!]{\linewidth}
    \centering
     \includegraphics[width=0.8\columnwidth]{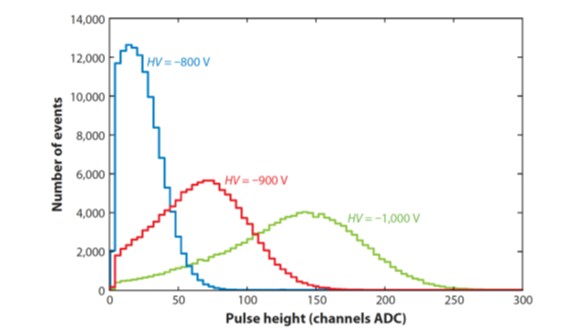}
     \includegraphics[width=0.75\columnwidth]{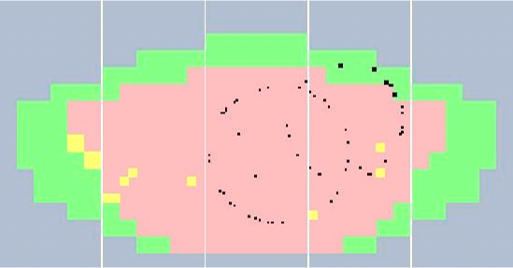}
\end{subfigure}
\caption{Multianode PMTs, performance: single photon pulse-height spectra (top)~\cite{mapmt1997}, a low multiplicity event as recorded in the HERA-B RICH detector (bottom)~\cite{hbrich}.}
\label{fig:hrich}
\end{figure}
Its use as a single photon sensor in particle physics experiments was pioneered in the HERA-B RICH~\cite{mapmt1997,hbrich}, and was later used in the COMPASS~\cite{compass-mapmt}, CLASS12~\cite{class12}, and GlueX~\cite{gluex} RICH detectors. Recently, it was installed in the upgraded RICH detectors of the LHCb experiment~\cite{lhcb-mapmt}; it is also planned for the RICH detector of the CBM experiment~\cite{cbm}.

\subsection{Micro Channel Plate PMT}

\begin{figure}[h!]
\centering
\begin{subfigure}[h!]{\linewidth}
    \centering
    \includegraphics[width=0.75\columnwidth]{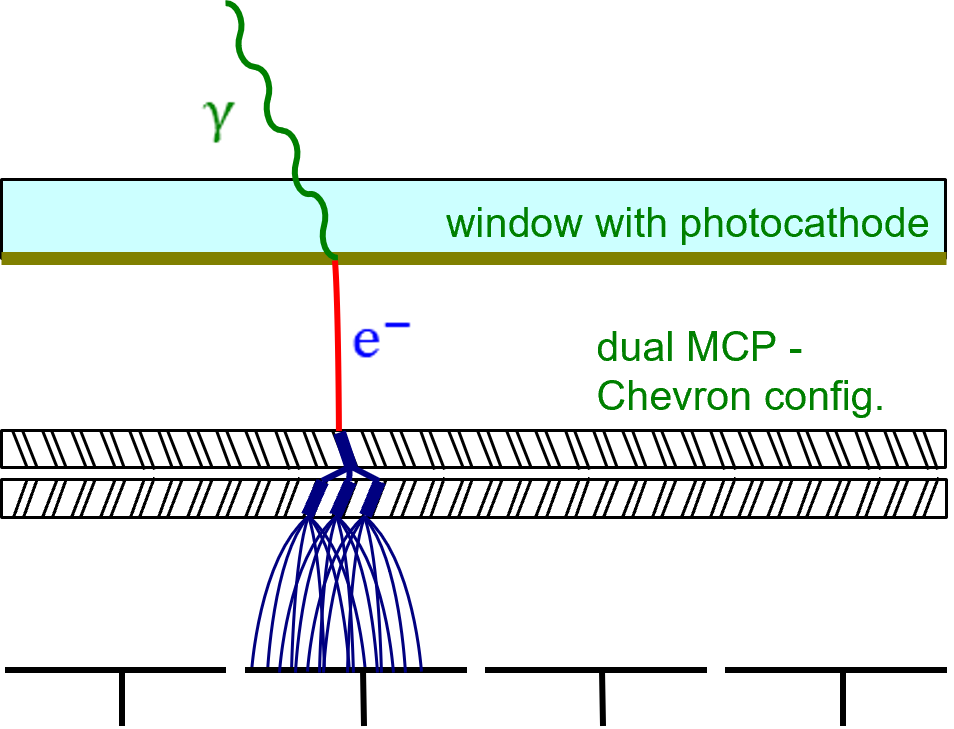}
    \includegraphics[width=0.75\columnwidth]{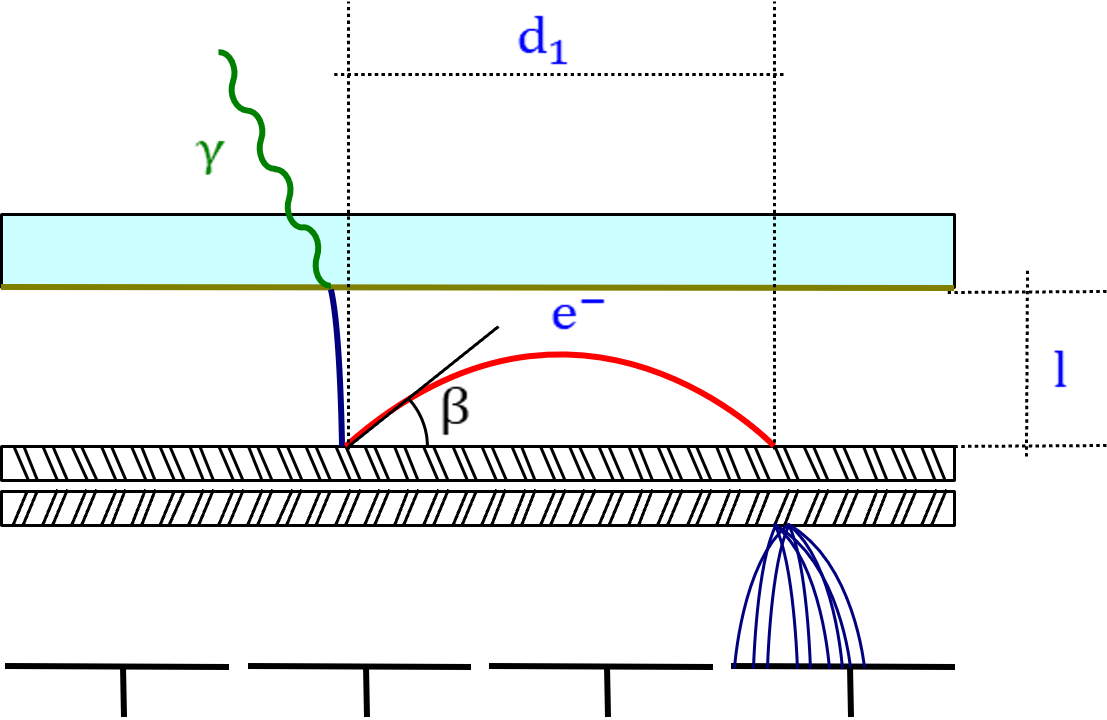}
\end{subfigure}
\caption{MCP-PMT: principle of operation (top), backscattered photoelectron (bottom)
.}
\label{fig:mcppmt-principle}
\end{figure}

In a Micro Channel Plate PMT (MCP-PMT), the discrete dynode multiplication chain is replaced by a continuous dynode, a microchannel plate, a thin glass plate with an array of microchannels, holes with a diameter of 10-100~$\mu$m. The gain of an MCP depends on the ratio of the microchannel length $L$ to its radius $R$, and amounts to typically 1000 for $L/R=40$, at an applied high voltage of around 1000~V.  In an MCP-PMT, two MCPs are usually employed in a chevron configuration, with channels slanted with respect to the plane of the MCP (Fig.~\ref{fig:mcppmt-principle}). 
The resulting overall gain typically amounts to $10^6$. The collection efficiency for photo-electrons is typically about 60\%. Since the MCP PMT is a thin device operating at high voltage, it has excellent timing properties, with a single photon resolution of 20-40~ps. Devices with microchannel radius $R \approx 10$~$\mu$m are immune to magnetic fields around 1~T pointing approximately perpendicularly to the sensor plane. 

These types of sensors are typically implemented as square tubes with 1-inch or 2-inch sides. An interesting new development is the Large Area Picosecond Photodetector (LAPPD), where its large size, 230~mm$\times$220~mm, makes it very attractive for covering large surfaces. 

MCP PMTs with a segmented anode are also position-sensitive. The anode is typically subdivided into $\approx 5 \times 5$~mm$^2$ large pads, but can also be segmented differently, depending on application needs. The anodes are either internal or capacitively coupled on the outside of the tube. Electrons produced in the multiplication process spread out when traveling from the second MCP  to the anode and can therefore induce the signal on more than one anode; the resulting charge sharing can be used to improve the spatial resolution. 
\begin{figure}[h!]
\centering
\begin{subfigure}[h!]{\linewidth}
    \centering
    \includegraphics[width=1.0\columnwidth]{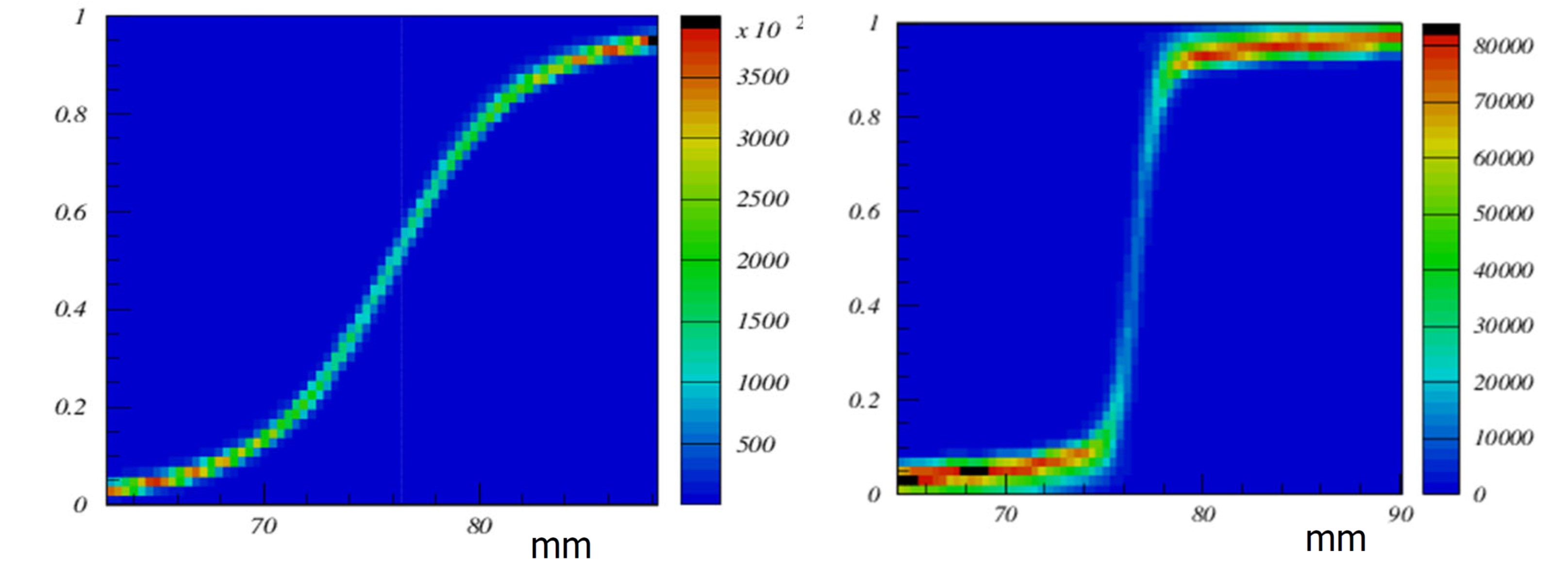}
\end{subfigure}
\caption{MCP-PMT: charge sharing in case of capacitive coupling (left, LAPPD Gen-II) and internal anodes (right, Photonis-Burle), fraction of the charge detected by the right pad as a function of a red laser spot position; both devices had the same pad size~\cite{korpar-burle-1,korpar-lappd}
.}
\label{fig:mcppmt-charge-sharing}
\end{figure}
As can be seen in Fig.~\ref{fig:mcppmt-charge-sharing}, charge sharing is more effective for capacitive coupling as the charge spreads over a larger area; since such an electrode is on the outside of the vacuum device, the pattern of read-out pads can be optimized for each specific application.  Finally, we note that the charge exiting the second MCP can also be detected in a pixelated ASIC like Timepix mounted within the vacuum housing~\cite{mcp-timepix}. 

\subsubsection*{Timing of MCP-PMTs}

\begin{figure}[h!]
\centering
\begin{subfigure}[h!]{\linewidth}
    \centering
    \includegraphics[width=0.85\columnwidth]{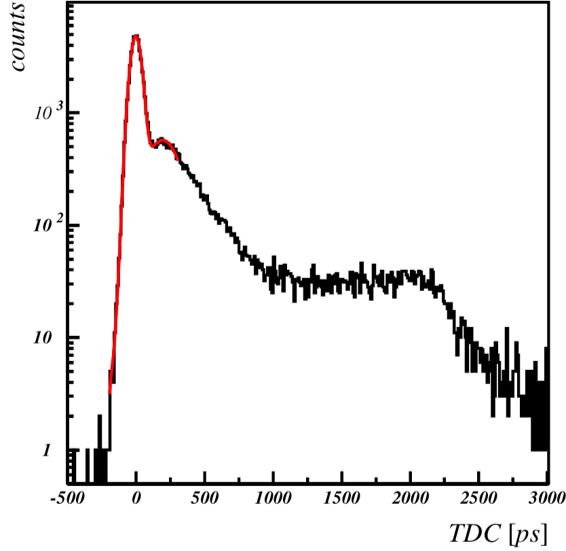}
\end{subfigure}
\caption{Single photon timing response of an MCP-PMT: a narrow main peak and tails, attributed to contributions from elastic (flat distribution) and inelastic back-scattering of photoelectrons~\cite{korpar-burle-1}
.}
\label{fig:mcppmt-timing}
\end{figure}

The typical single photon timing distribution (Fig.~\ref{fig:mcppmt-timing}) is characterized by a narrow main peak ($\sigma \approx 40$~ps) and considerably longer tails. The tails can be understood within a simple model~\cite{korpar-burle-1} and can be attributed to 
contributions from elastic (flat distribution) and inelastic back-scattering of photoelectrons (Fig.~\ref{fig:mcppmt-principle}). 

In addition to producing long tails in the timing distribution, photoelectron back-scattering reduces collection efficiency and gain and contributes to cross-talk in multi-anode PMTs
\footnote{The back-scattering contribution to cross-talk gets greatly reduced in the presence of magnetic field perpendicular to the tube window.}. In the simple model~\cite{korpar-burle-1}, the maximum range for backscattered photoelectrons is twice the distance between the photocathode and the first electrode, while the maximum delay is twice the photoelectron travel time to the first electrode; this model also predicts a flat arrival time distribution for the elastically scattered photoelectrons as seen in Fig.~\ref{fig:mcppmt-timing}. The experimental data confirm another prediction of this model: the tail width in the timing distribution can be significantly reduced by a decreased photocathode-MCP distance and by an increased voltage difference between the photocathode and the first electrode; the width of the prompt Gaussian peak decreases as an inverse square root of this voltage difference~\cite{korpar-lappd}. We also note that in cases where for timing only a single common output from the tube is required (e.g., if the MCP PMT is used as a time-of-flight detector), the electrode at the downstream end of the second MCP can serve the purpose~\cite{korpar-burle-2}.

\subsubsection*{Aging of MCP-PMTs}

\begin{figure}[h!]
\centering
\begin{subfigure}[h!]{\linewidth}
    \centering
    \includegraphics[width=0.85\columnwidth]{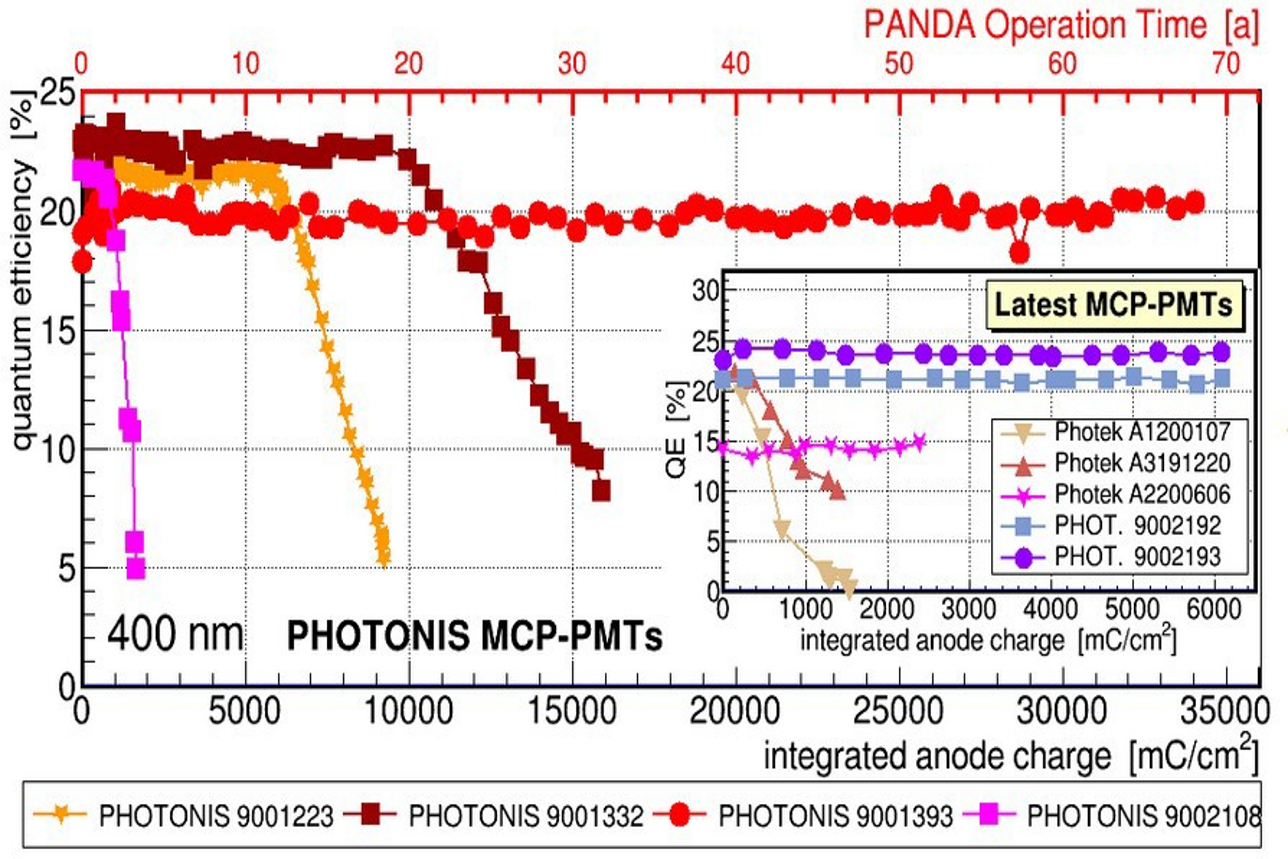}
    \includegraphics[width=0.85\columnwidth]{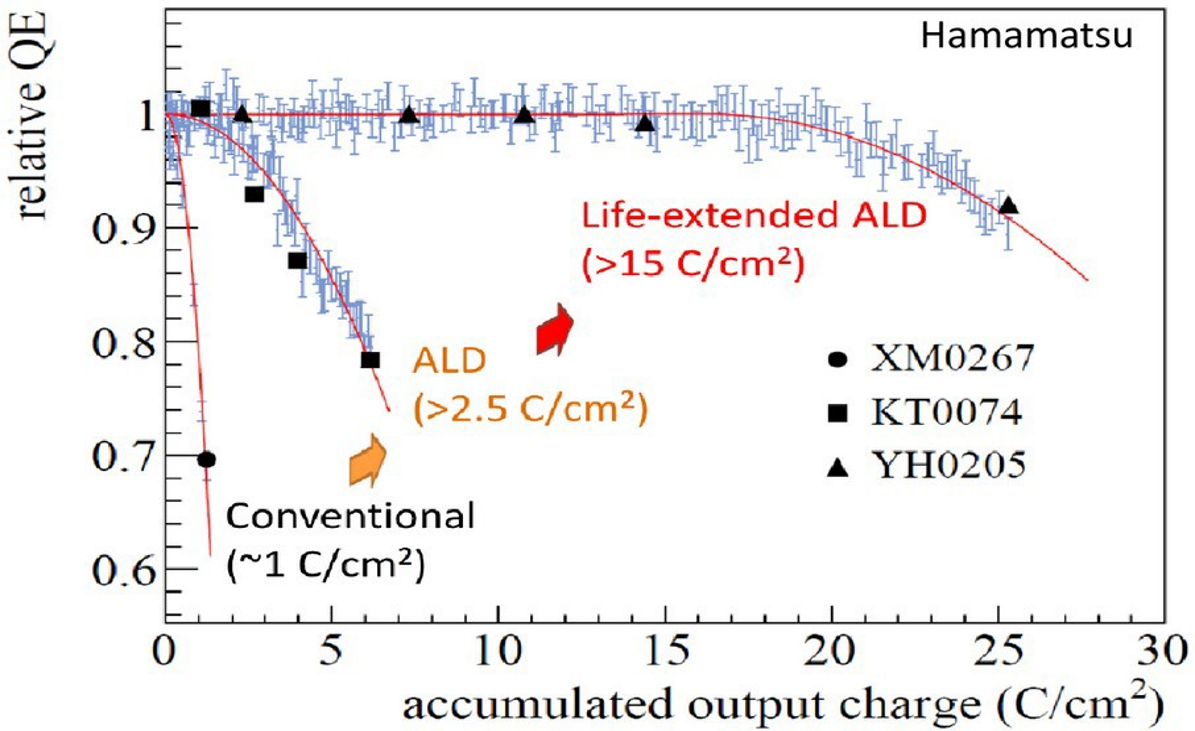}
\end{subfigure}
\caption{Aging of MCP-PMTs: quantum efficiency as a function of accumulated anode charge for Photonis~\cite{lehmann} and Hamamatsu MCP PMTs~\cite{inami-tf4}.}
\label{fig:mcppmt-aging}
\end{figure}

The main concern for the operation of MCP PMTs in high-intensity experiments is the photo-cathode degradation due to ions that are liberated in the impact of secondary electrons on the wall of microchannels. Recently a new method, atomic layer deposition (ALD) coating of MCP microchannel surfaces, was introduced to mitigate this effect. As can be seen in Fig.~\ref{fig:mcppmt-aging}, 
the MCP PMTs that have been treated with ALD show a two-orders of magnitude increase in photo-cathode lifetime. Hamamatsu 1-inch YH0205 devices tolerate a 20~C/cm$^2$ collected charge on the anode~\cite{inami-tf4}, and no QE degradation was observed for Photonis MCP-PMT (R2D2) up to 34~C/cm$^2$~\cite{lehmann}.

\subsection{Hybrid photodetectors}

\begin{figure}[h!]
\centering
\begin{subfigure}[h!]{\linewidth}
    \centering
    \includegraphics[width=0.95\columnwidth]{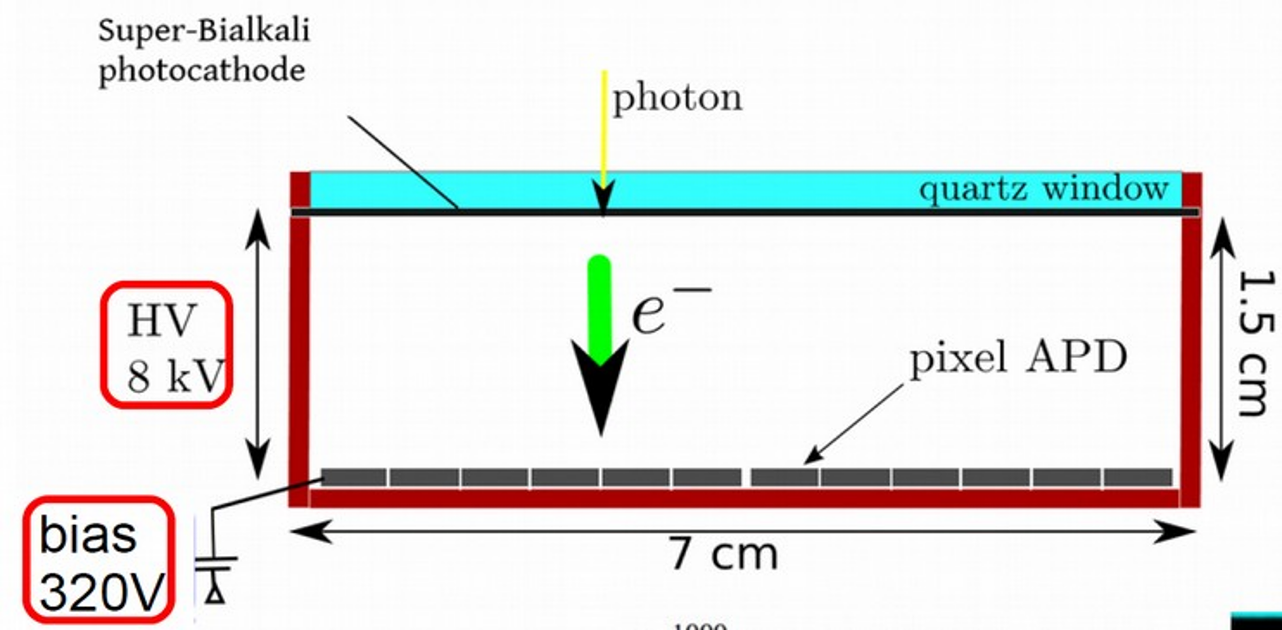}
\end{subfigure}
\caption{Hybrid photodetectors: principle of operation of an HAPD~\cite{hapd}.
}
\label{fig:hapd}
\end{figure}

In a hybrid photodetector (Fig.~\ref{fig:hapd}), photoelectrons are accelerated in a static electric field (over a voltage difference between 8~kV and 25~kV) and are then detected with a segmented PIN diode (known as HPD), an avalanche photodiode (HAPD), or with a silicon photomultiplier (VSiPMT). They found use in several large detectors, RICH1 and RICH2 of the LHCb detector in Run 1 and Run 2, Aerogel RICH detector of the Belle II spectrometer, and in the HCAL detector of the CMS experiment. 

Similarly as in the MCP-PMT, around 20\% of photoelectrons back-scatter from the first electrode (Si sensor in this case), and in the absence of magnetic field, their maximum range is twice the distance from the photocathode to the APD, corresponding to $\approx 40$~mm for the Belle II HAPD~\cite{hapd}. If the HAPD is placed in a magnetic field perpendicular to the entrance window, the scattered photoelectrons follow the magnetic field lines and get detected close to the original impact point, so that the kinetic energy of the photoelectron is deposited at the same pad it got scattered from~\cite{hapd}.

\section{Silicon photomultiplier - SiPM}

The only semiconductor sensor capable of detecting single photons is the silicon photomultiplier (SiPM), an array of APDs, avalanche photodiodes  (microcells or SPADs – single photon avalanche diodes) operated in the Geiger mode, i.e., above the APD breakdown  voltage.
\begin{figure}[h!]
\centering
\begin{subfigure}[h!]{\linewidth}
    \centering
    \includegraphics[width=0.85\columnwidth]{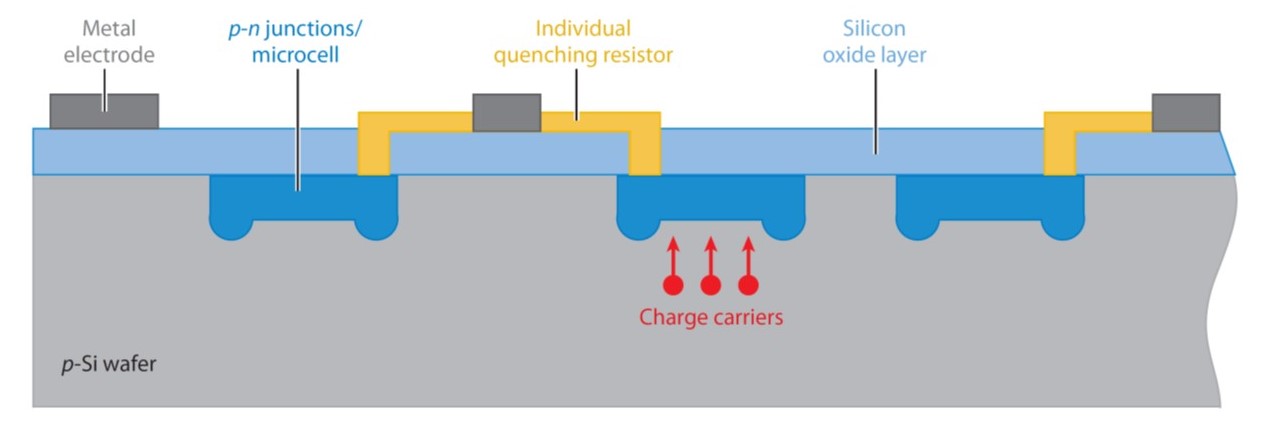}
\end{subfigure}
\caption{SiPM schematic~\cite{sspd-for-rich}
.}
\label{fig:sipm}
\end{figure}
In this sensor type, an absorbed photon generates an electron-hole pair (Fig.~\ref{fig:sipm}). An avalanche is triggered by the carrier in the high-field region resulting in a signal. The signal current results in a voltage drop at an external resistor, reducing the supply voltage below its breakdown value, and the avalanche is quenched (through passive or active quenching). Each triggered microcell contributes the same amount of charge to the signal.

The main challenge in using SiPMs for the detection of single photons is dark counts, produced by the thermal generation of carriers, trap-assisted tunneling, or band-gap tunneling~\cite{acerbi2019}. The pulse-height distribution of dark counts is namely the same as the single photon response. The typical dark-count rate dropped from $\approx 1$~MHz/mm$^2$ in early models to below 100kHz/mm$^2$  for more recent devices. The rates decrease at lower temperatures, they get roughly halved for every -8$^{\rm o}$C. Two more noise sources are the optical cross-talk, produced when photons emitted in an avalanche initiate a signal in a neighboring cell, and after-pulses, produced by trap-release of carriers or delayed arrival of an optically induced carrier in the same cell. 

The operation parameters of SiPM are correlated: a higher over-voltage (voltage in access of the breakdown voltage) and 
higher electric field in the sensor increase the probability of triggering an avalanche, thereby increasing the photon detection efficiency (PDE); it also results in faster signals and therefore better timing properties. At higher gain, the sensor also has a better signal-to-noise ratio. However, at a higher over-voltage, there is more optical cross-talk, a higher excess-noise factor (ENF), and there are more after-pulses - all this contributing to a possible deterioration of the fast timing properties~\cite{acerbi2019}. 

Special types of SiPMs are being developed to detect scintillation light of noble liquids where VUV sensitivity  (128~nm in LAr, 178~nm in LXe) is required together with operation at cryogenic temperatures. In this operation regime, SiPMs have very low dark count rates (below Hz/mm$^2$) dominated by band-band tunneling, reduced by low-field avalanche region, and an after-pulse rate $\approx 10$\%. In this case, a typical PDE value is about 20\%~\cite{retiere-vuv-sipm}.

\subsection{Fast timing of SiPMs}
\label{sec:sipm-timing}

\begin{figure}[h!]
\centering
\begin{subfigure}[h!]{\linewidth}
    \centering
    \includegraphics[width=0.8\columnwidth]{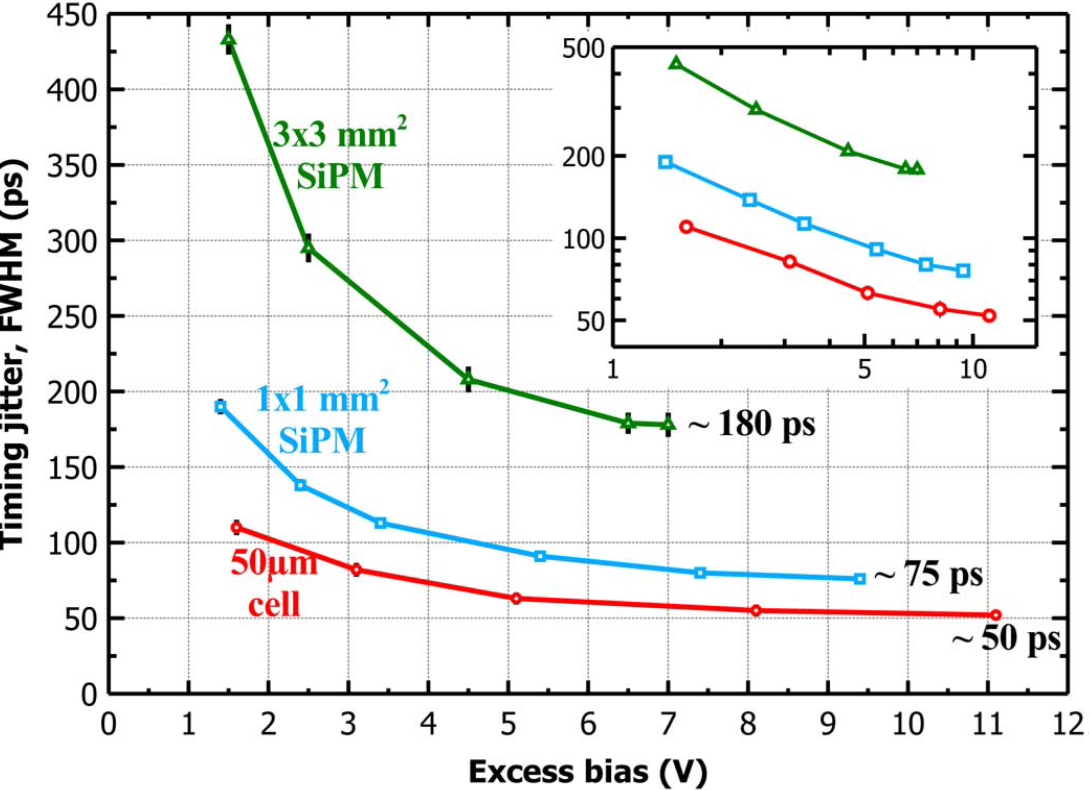}
    \includegraphics[width=0.8\columnwidth]{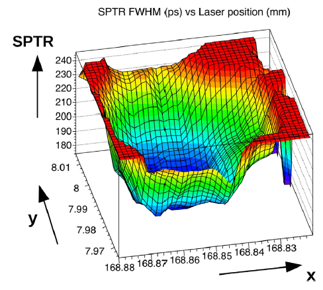}
\end{subfigure}
\caption{SiPM timing properties, top: variation of single-photon timing resolution (SPTR) as a function of the over-voltage for three different SiPM sizes with the same microcell~\cite{acerbi2015}; bottom: variation of single-photon timing resolution over the device surface~\cite{acerbi2019}.}
\label{fig:sipm-timing}
\end{figure}

The time resolution of SiPM microcells is excellent, with a $\sigma < 20$~ps, but deteriorates for larger devices. While the main contribution to this effect in the early types of SiPMs was the spread of the transient time across the device~\cite{korpar2015}, it is at present (Fig.~\ref{fig:sipm-timing}) due to a larger overall capacitance of larger devices (and the corresponding reduced signal slope)~\cite{acerbi2015}
as well as due to the non-uniformity within a microcell, in particular at its edges~\cite{korpar2015,acerbi2019}. 
The latter effect can be mitigated by masking of outer regions of microcells with worse time resolution~\cite{fbk-nemallapudi}. 

\begin{figure}[h!]
\centering
\begin{subfigure}[h!]{\linewidth}
    \centering
    \includegraphics[width=0.85\columnwidth]{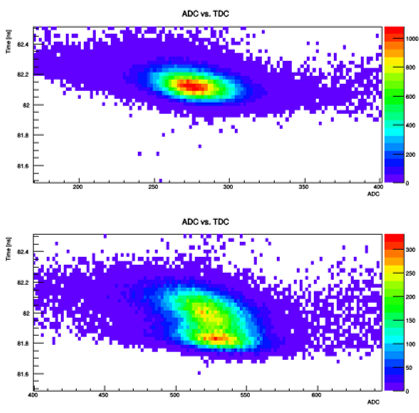}
\end{subfigure}
\caption{Timing of events with single-cell and two-cell hits, timing vs amplitude for  AdvanSiD SiPM ASD-NUV3S-P-40~\cite{dolenec2017}.}
\label{fig:double-hits}
\end{figure}

In timing properties of multi-cell signals we note that the optical cross-talk contribution to multi-cell signals spoils the timing distribution such that the resolution does not scale with $1/N^{1/2}$ for $N$ microcell hits~\cite{acerbi2015,dolenec2017}. In the case of two hits, there are events with two-photon hits and events with a combination of a single photon hit with optical cross-talk, with timing between single and double microcell signals as illustrated in Fig.~\ref{fig:double-hits}. 
The ratio between the two contributions changes with light intensity confirming the optical cross-talk origin. Clearly, in the case of more than two hits, there are even more components~\cite{acerbi2015}.

A reduction of the optical cross-talk involves stopping photons emitted in the Geiger discharge from reaching one of the neighboring microcells. Recently, a novel technology was developed where metal-filled deep trench isolation is employed to strongly suppress optical crosstalk~\cite{GolaRICH2022}. Other possible modifications include a low electric field design and a layout optimized for timing.



\subsection{Light concentrators} 

\begin{figure}[h!]
\centering
\begin{subfigure}[h!]{\linewidth}
    \centering
     \includegraphics[width=0.65\columnwidth]{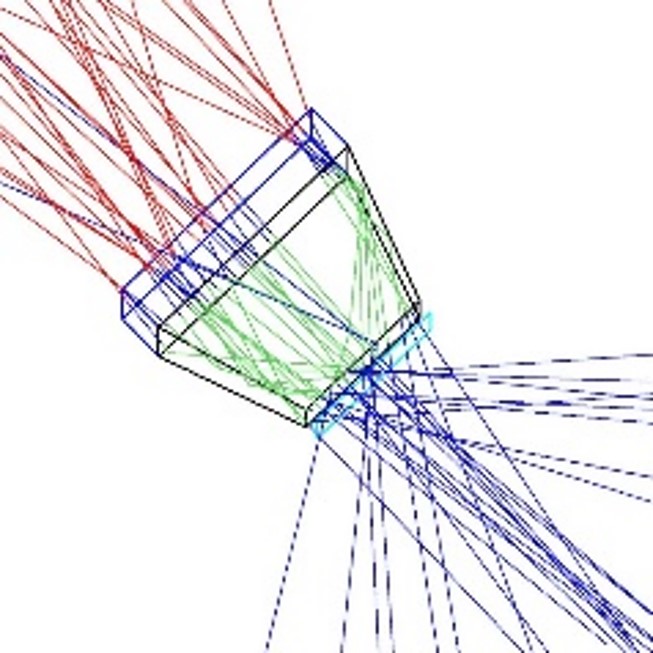}
     \includegraphics[width=0.65\columnwidth]{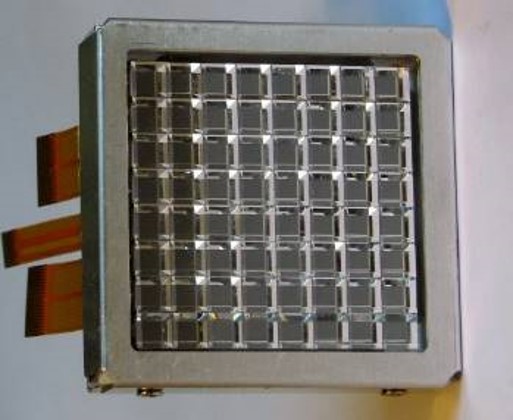}
\end{subfigure}
\caption{Light concentrator in front of a SiPM (top), bottom: an array of light concentrators coupled to an array of SiPMs (Hamamatsu 64 channel MPPC module S11834-3388DF)~\cite{dino2015}.}
\label{fig:sipm-lc-1}
\end{figure}

Light concentrators can be used with SiPMs in two ways. When used at the device level with lenses or Winston cones (Fig.~\ref{fig:sipm-lc-1}), 
they allow reducing the area of the SiPM while keeping the active area of the photon detector. This results in a decrease of the dark-count rate per unit area of the photon detector as well as in reducing the cost of the detector. It also enables the use of smaller, and therefore faster devices, as discussed in Sec.~\ref{sec:sipm-timing}. An example of an array of light concentrators coupled to an array of $3 \times3 $~mm$^2$ SiPMs is shown in Fig.~\ref{fig:sipm-lc-1}; this combination was successfully tested in a prototype RICH beam test~\cite{dino2015}. 

\begin{figure}[h!]
\centering
\begin{subfigure}[h!]{\linewidth}
    \centering
     \includegraphics[width=0.85\columnwidth]{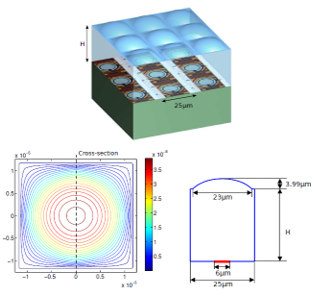}
     \includegraphics[width=0.85\columnwidth]{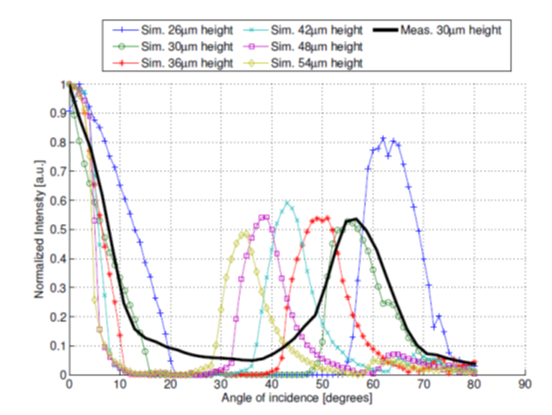}
\end{subfigure}
\caption{Microlenses in front of SiPM cells: principle (top), angular acceptance  as a function of a geometrical parameter (height of the array) for a 128x128 array of 6~$\mu$m diameter CMOS SPADs at 25~$\mu$m pitch and a 5\% fill factor~\cite{pavia2014}.}
\label{fig:microlens}
\end{figure}

Light concentration at the microcell level (micro-lenses, diffractive lenses, meta lenses) can be used to compensate for a low fill factor, as well as to concentrate light in the microcell center to improve the timing~\cite{charbon2017}. In Fig.~\ref{fig:microlens} 
the principle of microlensing is shown together with the dependence of the angular acceptance for different geometrical parameters of the microlensing array for a micro-lens array coupled to a CMOS SPAD array~\cite{pavia2014}. The microlense arrays can be implemented in a multi-step process with photoresist masters and UV-curable hybrid polymers~\cite{bruschini-microlenses}.

As illustrated in Fig.~\ref{fig:microlens}, for both types of light concentrators a higher light concentration results in a narrower angular acceptance~\cite{pavia2014}, following Liouville's theorem. 
We also note that in imaging light concentrators the impact angles on the sensor are smaller; they can also be used with position-sensitive arrays. In non-imaging light concentrators,  the impact angles on the sensor are larger, so that they are preferably directly coupled to the sensor.
We finally note that for an efficient light collection from scintillator crystals, novel materials are being considered like photonic crystals and metamaterials~\cite{lecoq-photonic-c-2010,light-extraction}.

\subsection{Radiation damage of SiPMs}

The rate of dark counts, the main challenge in using SiPMs for detection of single photons, increases linearly with neutron fluence due to bulk damage.
\begin{figure}[h!]
\centering
\begin{subfigure}[h!]{\linewidth}
    \centering
    \includegraphics[width=0.85\columnwidth]{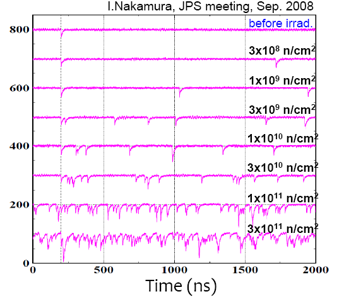}
\end{subfigure}
\caption{Dark count rate of irradiated SiPMs~\cite{nakamura2007}.}
\label{fig:sipm-neutrons}
\end{figure}
As can be seen from Fig.~\ref{fig:sipm-neutrons}, this becomes a show-stopper at neutron fluences above $\approx 10^{11}$n~cm$^{-2}$  in applications where single (or few) photon sensitivity is required
\footnote{E.g., the expected fluence in the most exposed part of the ARICH area of Belle II is   $2 \times 10^{12}$n~cm$^{-2}$ }.

Possible mitigation strategies include the use of wave-form sampling readout electronics, operation of SiPMs at lower temperatures, periodic annealing at elevated temperatures, reducing recovery time to lower cell occupancy, and radiation-resistant SiPMs, possibly produced from other materials.

It has been shown that beyond fluences of $10^{7}-10^{8}$n~cm$^{-2}$  there is little correlation between the dark count rate before and after the irradiation, and sensors produced with different technologies seem to behave in a similar way. 

\begin{figure}[h!]
\centering
\begin{subfigure}[h!]{\linewidth}
    \centering
    \includegraphics[width=0.85\columnwidth]{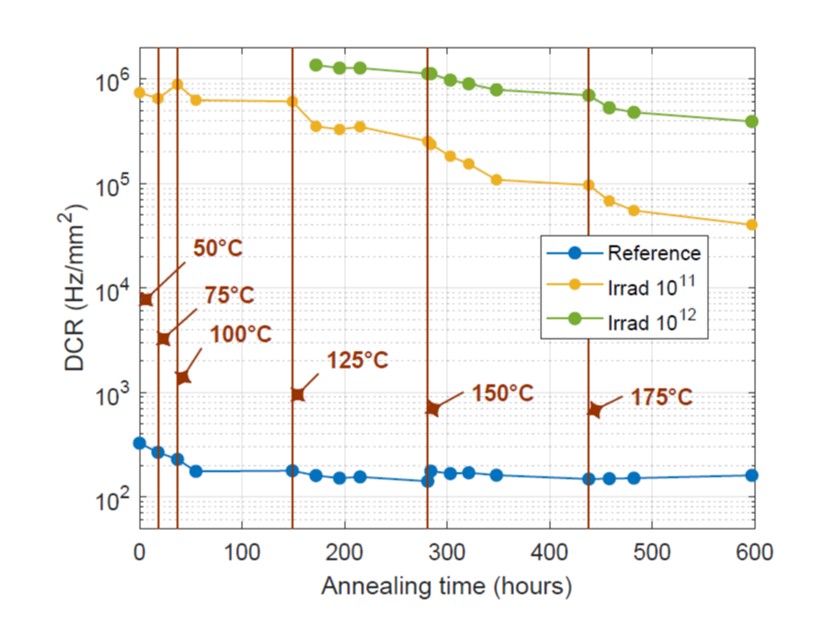}
\end{subfigure}
\caption{Annealing of irradiated SiPMs as a function of annealing time: comparison non-irradiated (blue) and irradiated up to $10^{11}$n~cm$^{-2}$ (yellow), $10^{12}$n~cm$^{-2}$ (green) and $10^{13}$n~cm$^{-2}$ (orange)~\cite{calvi2019}.}
\label{fig:annealing}
\end{figure}

Annealing at elevated temperatures seems to be one of the most promising mitigation measures. While room temperature annealing for samples showed little effect, several studies have reported promising results when annealing at elevated temperatures~\cite{calvi2019}. In Fig.~\ref{fig:annealing}, 
dark counts are shown for operation at -30$^{\rm o}$C of irradiated Hamamatsu S13360-1350CS SiPMs as a function of annealing time and temperature~\cite{calvi2019}. In the same study, it was also shown that annealing helps even when operating at  77~K~\cite{calvi2019}.

One of the possible solutions could also be the use of new wider band-gap materials with a possibly lower rate of thermally-generated dark counts, higher radiation resistance, possible operation at higher temperatures, and (V)UV sensitivity. The dark count rate is in this case dominated by trap-assisted tunneling. An interesting material is 4H-SiC with a gap of $E_g=3.26$~eV. At present, a PDE of about 10\% was achieved, with a dark count rate of above 1~MHz/mm$^2$ and a  nonuniform response~\cite{HSiC}. We note here that the early versions of SiPMs also had similar properties so it is expected that this will improve with more experience in the production. 

\section{Gas based photon detectors}

In gas-based photon detectors, one of the electrodes of the gas chamber is covered by a solid photo-sensitive substance. The most common material is CsI evaporated on one of the cathodes.  The largest scale application of this concept is the ALICE RICH with 11~m$^2$ of photosensitive multiwire chambers~\cite{alice-rich}. Instead of a multiwire chamber detectors with multiple GEM or thick GEM (THGEM) gas amplification stages with transmissive or reflective photocathodes can be used. An example of such a combination is the photon detector for the RICH counter of the COMPASS experiment with the amplification stage comprised of a pair of THGEM steps and a MicroMegas~\cite{DallaTorre2020}.

For operation in environments with high particle flux densities over extended periods of time, aging due to photocathode damage by ions produced in avalanches can become a limiting factor. One of the mitigation strategies is to block the drifting of ions towards the photocathode by, e.g., non-aligned holes in the consecutive GEM gas amplification stages.   

Recent developments include detectors with smaller pads, as well as novel photocathode materials, like a nano-diamond layer.

\section{Summary and outlook}
\label{sum}
Next generation of experiments in particle physics will require position-sensitive low-light-level sensors with faster timing, a wider spectral range, and an improved radiation tolerance. Many new interesting developments in this research of light sensors are underway, in particular for SiPMs and MCP-PMTs; unfortunately, because of space limitations not all of them could be covered in this paper. 

We finally note that a detector R\&D collaboration (DRD4) is being set up to facilitate collaboration in this area of research; the collaboration is expected to formally start its activities early in 2024.

\section{Acknowledgements}

I thank S.~Korpar, R.~Pestotnik, R.~Dolenec, A.~Seljak, N.~Harnew, J.~Milne, C.~Bruschini, A.~Gola for useful discussions in the preparation of the review.  
This work was supported by the following funding sources: European Research Council, Horizon 2020 ERC-Advanced Grant No. 884719, and Slovenian Research and Innovation Agency grants No. J1-9124, J1-4358 and P1-0135.

\end{document}